# Topological properties and shape of proliferative and non-proliferative cell monolayers

Daria S. Roshal,[1*] Karim Azzag,[2*] Kirill K. Fedorenko,[1] Sergei B. Rochal,[1**] Stephen Baghdiguian.[3***]

[1]Physics Faculty, Southern Federal University, Rostov-on-Don, Russia.
[2]Lillehei Heart Institute, Department of Medicine, University of Minnesota, Minneapolis, MN, USA
[3]Institut des Sciences de l'Evolution-Montpellier, Université de Montpellier, CNRS, Ecole Pratique des Hautes Etudes, Institut de Recherche pour le Développement, Montpellier, 34095 France.
*Equal contribution
**E-mail: rochal_s@yahoo.fr
***E-mail: stephen.baghdiguian@umontpellier.fr

**Abstract**

During embryonic development, structures with complex geometry can emerge from planar epithelial monolayers and to study these shape transitions is of key importance for revealing the biophysical laws involved in the morphogenesis of biological systems. Here, using the example of normal proliferative monkey kidney (COS) cell monolayers, we investigate global and local topological characteristics of this model system in dependence on its shape. The obtained distributions of cells by their valence demonstrate a previously undetected difference between the spherical and planar monolayers. In addition, in both spherical and planar monolayers, the probability to observe a pair of neighboring cells with certain valences depends on the topological charge of the pair. The zero topological charge of the cell pair can increase the probability for the cells to be the nearest neighbors. We then test and confirm that analogous relationships take place in a more ordered spherical system with a larger fraction of 6-valent cells, namely in the non-proliferative epithelium (follicular system) of ascidian species oocytes. However, unlike spherical COS cell monolayers, ascidian monolayers are prone to non-random agglomeration of 6-valent cells and have linear topological defects called scars and pleats. The reasons for this difference in morphology are discussed. The morphological peculiarities found are compared with predictions of widely used vertex model of epithelium.

## I. INTRODUCTION

Nowadays, it is well known that along with genetic control, physical and geometrical laws drive the formation and properties of biological systems like viruses or epithelia. However, the path to this knowledge began a long time ago. Back in 1665, Robert Hooke revealed that the boundaries of cells look like polygons. The first statistical analysis of the polygon distribution by side number in the epidermis of a cucumber was made in 1926 [1]. Much later, in 2006, it was demonstrated that similar distributions are typical of several proliferative epithelia and a hypothesis of the distribution invariance, or topological invariance, was formulated [2-5]. Then, the hypothesis was confirmed for many plant and animal proliferative epithelia [6-8] and it became clear that the topological invariance is closely related to the physiological invariance of normal epithelium [9-10]: in all eumetazoans, the epithelium acts as a selective paracellular barrier that controls fluxes of nutrients, regulates ion and water balance and limits host contact with antigens and microbes. Recently, topological analysis revealed common morphological features (specific topological defects) of the spherical postmitotic follicular system of ascidian eggs [11-13] and colloidal crystals [14]. In this context, it is interesting to note that other topological defects like those typical of liquid crystals control apoptosis and extrusion of cells from the monolayer [15].

Topological methods have been successfully used to rationalize various morphological transformations possible in the epithelium. For instance, normal development of epithelium involves mesenchymal-to-epithelial transition leading to the dramatic decrease of cell mobility and mitotic rate. At this transition, cells take more regular shapes with smaller average perimeter [16]. The reverse process, leading to the appearance of elongated cells with high mobility, is called the epithelial-to-mesenchymal transition (EMT) [17] and cells undergoing oncogenic EMT drive metastasis process [18]. Epithelial-mesenchymal hybrid state in cancer progression is also studied [19]. A recent article [20] shows that specific topological defects appear at a brittle-to-ductile transition induced by a competition between the cell rearrangement and cell detachment.

Equally or even more exciting are the shape transitions occurring in epithelia and changing their curvature. During the early stages of embryonic development in mammals, planar epithelial monolayers can form structures of complex geometry. As an example, the primitive endoderm transforms into the yolk sac, which has a variety of functions [21,22] including a major role in the establishment of the anteroposterior axis of the embryo [23]. Shape transition also occurs during the formation of epithelial tubes which are fundamental structures across the metazoan phyla and provide an essential functional component of many of the major organs [24]. The buckling observed during the formation of intestinal villi is another example of the transformation from the flat structure [25,26].

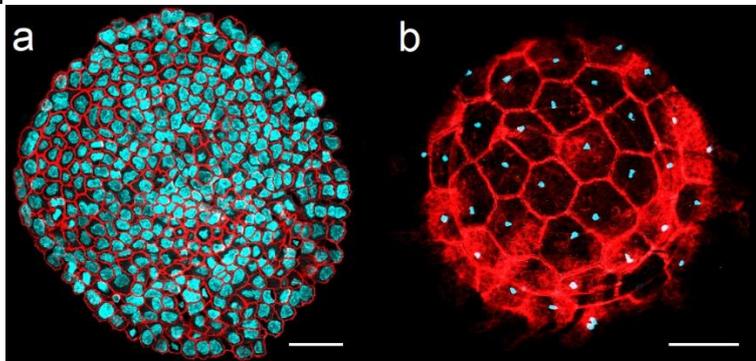

FIG. 1. Structural organization of COS cell organoid (a) and epithelial follicular cells of *Ciona intestinalis* egg (b) (ascidian species). Cell contours (*i. e.* cortical actin) are shown in red; cell nuclei are cyan (see Appendixes A, B, and C for more details of culture, collection, and labeling). Scale bars are 100 and 50 μm for panels (a, b), respectively.

The aim of this article is to analyze the relationships between the local topological characteristics of individual cells, shape of epithelial structures and degree of their order. The rest of the study is organized as follows. In the next section we present mathematical properties of spherical and planar cell monolayers and give the definition of non-random pair correlations characterizing the valences of the nearest cells. In section III we investigate and compare the COS cell monolayers (monkey kidney cells immortalized by the SV40 virus T antigen) with planar and spherical shapes. The growth of COS cell monolayers occurs by intercalated cell division. The various shapes of the monolayers are obtained due to a simple modification of the cell culture conditions [13] (see also Appendix A). In several dozens of flat and spherical samples (see Fig. 1a), we analyze the fractions of cells with different numbers of nearest neighbors and studied patterns in the arrangement of pairs of neighboring cells depending on their valency. In section IV, to expand our study and test the robustness of the theoretical approach, we perform a topological analysis of a non-proliferative follicular epithelium (see Fig. 1b) covering oocytes of solitary ascidians (see Appendix B). In contrast to COS cells, morphogenesis of this spherical epithelium occurs by cell accretion [13]. Then, in section V, the discovered structural peculiarities are compared with predictions of a widely used vertex model [27-30]. The last section VI is devoted to discussion and conclusion.

## II. SOME TOPOLOGICAL AND PROBABILISTIC PROPERTIES OF PLANAR AND SPHERICAL CELL MONOLAYERS

Before performing the structural characterization of cell monolayers, we recall some general topological properties of these packings. Cell boundaries represent polygons that are usually convex and form a tessellation of the monolayer surface, which can be characterized by a distribution of polygon types (DOPT), in other words, by fractions of cells with different valences [2]. The considered tessellation is similar to the Voronoi [31] one, which, in turn, is a dual of Delaunay triangulation [32].

In addition to DOPT, both planar and spherical monolayers can be characterized by the value

$$\Delta = \sum_i P_i \cdot (6 - i), \qquad (1)$$

where $P_i = N_i/\Sigma_j N_j$ is the fraction of $i$-valent cells and $N_j$ is the number of $j$-valent cells. The quantity $Q = \sum_i N_i \cdot (6 - i)$, proportional to $\Delta$, is called the topological charge [11,33,34] and characterizes a whole system; the topological charge $q$ of each $i$-valent cell is $6 - i$. For an infinite planar monolayer $Q = 0$ and, consequently, $\Delta = 0$ [10]. For any triangulated structure with spherical geometry, $Q = 12$ and $\Delta = 12/N_{tot}$, where $N_{tot}$ is the total number of cells in the structure. In an infinite planar monolayer, the DOPT is balanced: the number of $n$-valent cells with $n < 6$ balances the number of $n$-valent ones with $n > 6$. In this equilibrium, the weights of 5- and 7-valent cells (see Eq. (1)) are equal to one, the weights of 4- and 8-gons are twice as large, etc. Note that the exact values of $\Delta$ for plane and spherical geometries may be violated by the finite dimensions of the sample. In this case, the value of $\Delta$ deviates randomly from zero for finite flat samples and from $12/N_{tot}$ for fragments of spherical samples.

In any triangulation of an infinite plane, the average number of lines outgoing from one node is six, and the most topologically perfect triangulation is that in which all nodes are 6-valent. However, even the most perfect spherical triangulation must have at least twelve 5-valent vertices. Previously, we proposed [11] to characterize the defectiveness of monolayers with spherical geometry by the absolute value of total negative charge of all topological defects $Q_- = \sum_{i \geq 7} N_i \cdot (i - 6)$, where $N_i$ is the number of cells (particles) having $i$ neighbors, $(6 - i)$ is the topological charge of each cell or particle. In the most topologically perfect spherical structures, this value is exactly 0. Along with $Q_-$ one can introduce the relative topological defectiveness $\tilde{Q} = Q_-/N_{tot}$. The latter value is easily calculated as $\tilde{Q} = \sum_{i \geq 7} P_i \cdot (i - 6)$.

Before proceeding to a mathematical definition of pair correlations, we note that due to visualization specifics (see Appendix C), the cell nuclei in the micrographs obtained are always clearly visible, while the cell boundaries are often indistinguishable. Therefore, to analyze the images, we use the Voronoi tessellation and Delaunay triangulation with the nodes located at the centers of the nuclei. Thus, strictly speaking, we use the valency of cell nuclei instead of the valency of cells; the same approach was used in our previous works and the equivalence between two methods was discussed there as well [10-11].

When the triangulation edges are found, we can introduce the pair correlation functions that are usually applied for abiotic partially disordered systems [34,35]. Let us denote the total number of triangulation edges as $L_{tot}$, and the number of triangulation edges connecting cells with $i$ and $j$ neighbors as $L_{ij}$. Then the pair correlation function $C_{ij}$ is the fraction of all edges connecting the cells with the specified numbers of neighbors $C_{ij} = L_{ij}/L_{tot}$. The same definition can also be used for experimentally observed structures, however, an additional consideration of a way to calculate correctly $L_{tot}$ and $L_{ij}$ is needed (see Appendix D). Note that the introduced pair correlation functions do not vanish even in random structures, and we need to define a nonrandom part of the correlations.

To do this, we consider a hypothetical (random) triangulated structure that is characterized by a certain DOPT and contains $N$ triangulation nodes, where $N \to \infty$. Since each triangulation edge has two ends, the number of ends related to all $k$-valent cells is equal to $NP_k k$. Then the total number of ends is equal to $\sum_k (NP_k k)$, while the fraction of ends related to $i$-valent cells reads:

$$P_i^{end} = \frac{iP_i}{\sum_k (P_k k)}. \qquad (2)$$

The probability that a pair of independent events will occur is equal to the product of the probabilities of these events. Therefore, in the hypothetical random structure introduced by us, the fraction $P_{ij}^{rnd}$ of triangulation edges connecting *i*-valent and *j*-valent cells (where $i \neq j$) is proportional to a product between two fractions (2):

$$P_{ij}^{rnd} = \frac{2ijP_iP_j}{[\sum_k(P_k k)]^2}, \quad (3)$$

where coefficient 2 arises from the equivalence of triangulation edge ends. For cells with the same valence, the coefficient 2 disappears:

$$P_{ii}^{rnd} = \frac{i^2 P_i^2}{[\sum_k(P_k k)]^2}. \quad (4)$$

Note that probabilities (3-4) satisfy the normalization $\sum_{i,j} P_{ij}^{rnd} = 1$, and for an infinite flat or finite spherical monolayer the denominators of Eqs. (3,4) can be simplified. Using the identity

$$\sum_k(P_k k) = 6 + \sum_k[P_k \cdot (k-6)]$$

and the properties of $\Delta$ [Eq. (1)], one can see that in the flat case the sum in the denominator is equal to 6, while in the case of spherical geometry this sum equals $6 - 12/N$.

Thus, we can introduce the nonrandom contribution $C'_{ij}$ to the pair correlation function $C_{ij}$:

$$C'_{ij} = C_{ij} - P_{ij}^{rnd}, \quad (5)$$

where for small structural fragments, the probabilities (3-4) should be calculated directly, without denominator simplifications (valid in the limit of $N \to \infty$). In addition, we note that the positive or negative sign of the value (5) means that the cells of the valences *i* and *j* tend or do not tend to locate near each other (see Appendix E).

## III. MICROGRAPH ANALYSIS OF COS CELL MONOLAYERS WITH PLANAR AND SPHERICAL GEOMETRY

COS are fibroblast-like cell lines derived from monkey kidney tissue. COS cells were obtained by immortalizing CV-1 cells with a version of the SV40 virus that can produce large T antigen but has a defect in genomic replication. The CV-1 cell line in turn was derived from kidney of the African green monkey. The acronym "COS" means that the cells are CV-1 (simian) in Origin and carry the SV40 genetic material. The growth and visualization of flat and spherical COS cell monolayers are described in Appendices A and C.

The morphological characterization of the COS monolayers (see Fig. 2) was carried out using 33 and 26 images of flat and spherical structures, respectively. The analyzed areas in the flat monolayers contained from 73 to 758 cells, while in the spherical case they included from 17 to 399 cells (on the visible hemisphere). Note that we discard the cells located too close to the image border, for which the number of neighbors cannot be determined (see Appendix D for details). Thus, DOPT and pair correlations were calculated for each image. All averaging procedures over images were performed with the weighting factor, which is the number of Voronoi cells in the image. The procedures for spherical and flat samples were performed separately.

Fig. 2(a) shows a typical flat COS monolayer with 98 clearly visible nuclei. The Delaunay triangulation, the nodes of which are the centers of the nuclei, is superimposed on the monolayer. 53 nuclei that are far from the borders of the image are colored according to their valences. For the rest of the nuclei marked with grey circles in their centers, it is impossible to reliably determine the number of neighbors. Fig. 2(b) shows the Voronoi tessellation for the same monolayer. The 53 colored Voronoi cells correspond to the nuclei with reliably determined valences. The topology of a typical spherical monolayer (containing 73 COS cells in the visible region) is analyzed in Fig. 2(c,d). The coloring of nuclei in Fig. 2(c) matches one in Fig. 2(a). Fig. 2(d) shows the Voronoi cells for 51 monolayer nuclei for which the number of neighbors is correctly determined.

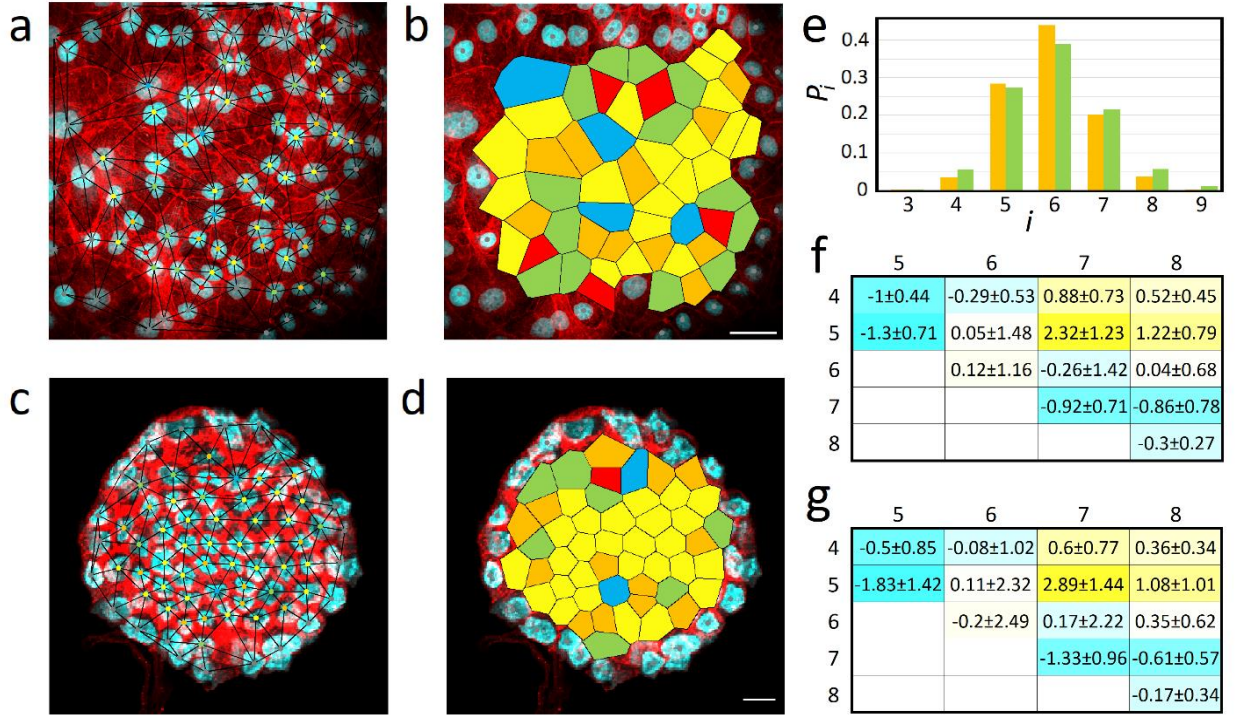

FIG. 2. Characterization of African green monkey (COS) kidney cells. (a,b) Typical image of a flat cell monolayer with superimposed Delaunay triangulation and Voronoi tessellation. (c,d) Typical sample of a spherical cell monolayer and its graphical analysis. Scale bars (15 μm) are shown with white lines. The triangulation nodes are the centers of cell nuclei. Red, orange, yellow, green, and blue nodes correspond to cells with 4,5,6,7,8 neighbors. Voronoi polygons are colored accordingly. Grey nodes lie at the centers of nuclei located too close to the image boundary; it is impossible to accurately determine the number of their neighbors (see Appendix D). (e) Distribution of cells according to the number of their sides for averaged data on flat and spherical monolayers. Orange and green colors correspond to spherical and flat monolayers, respectively. (f,g) Averaged values and their standard deviations for planar (f) and spherical (g) $C'_{ij}$ functions, respectively. Values $C'_{44} \pm \Delta C'_{ij}$ for planar and spherical case are $(-0.13 \pm 0.12) \cdot 10^{-2}$ and $(-0.07 \pm 0.15) \cdot 10^{-2}$, respectively. All the values presented in two tables should also be multiplied by $10^{-2}$. Table cells with positive and negative values of $C'_{ij}$ are highlighted in shades of yellow and blue, respectively. The color brightness is proportional to the value of $C'_{ij}$

DOPTs for averaged data on flat and spherical monolayers are shown in Fig. 2(e). These distributions are typical for other proliferative epithelial monolayers [2]. Indeed, for both types of monolayers the value of $P_6$, which mainly determines the DOPT [2,10], fall within the conventional range from 0.38 to 0.48. The values of $\Delta$ calculated for histograms shown in Fig. 2(e) are 0.02 and 0.06 for planar and spherical cases, respectively. Since $\Delta$ is a linear combination of probabilities $P_i$ and this value should vanish in an infinite planar monolayer, its difference from zero characterizes a statistical error in $P_i$ determination for the planar case. It is reasonable to assume that the error for spherical fragments is close. Then, the threefold increase in the value of $\Delta$ for the spherical case can only be explained by the presence of Gaussian curvature. Thus, the difference between geometries of spherical and planar monolayers manifests itself in the difference between DOPTs.

Averaged correlations $C'_{ij}$ calculated using Eq. (5) and corresponding standard deviations $C'_{ij}$ are shown in Fig. 2(f,g). In both tables corresponding to planar and spherical cases, the correlations $C'_{57}$ have positive values which are substantially larger than the standard deviations. All other correlations are small and close in magnitude to the standard deviation. It is pointless to discuss the magnitude of the latter correlations, but we can talk about their sign: in both cases the values of $C'_{47}$, $C'_{58}$ & $C'_{48}$ are positive, the values of $C'_{55}$, $C'_{45}$, $C'_{77}$ & $C'_{78}$ are negative and other correlation values vanish. Thus, according to our results in both cases the pair correlations are qualitatively similar and cannot be associated with the monolayer curvature. Only the value of $\Delta$ distinguishes clearly between spherical and planar COS monolayers.

COS monolayers are conventional proliferative monolayers with the typical DOPT. In the next section, we discuss what does the proposed approach reveal in the design of post-proliferative spherical monolayers corresponding to follicular system of ascidian eggs. These biological packings are of great interest because, depending on the species of ascidians, they are characterized by the value $P_6$ up to 0.7 and have topological scars and pleats [11, 36], which also indicate their relatively greater structural ordering. As we demonstrate below, in these spherical monolayers the total topological charge of a cell pair analogously controls the sign of nonrandom pair correlations. In addition, a significant positive nonrandom correlation emerges between the topologically neutral (6-valent) nearest cells.

### IV. MORPHOLOGY OF EPITHELIAL MONOLAYERS OF ASCIDIANS EGGS

Below, we analyze 140 samples of epithelial monolayers with spherical geometry (EMSG) covering eggs of 8 species of ascidians (see Appendix A). In our work [11], where the same monolayers were presented, the analogies between the least defective ascidian EMSG and spherical colloidal crystals have been demonstrated. It was shown that linear topological defects called scars and pleats (chains with alternating 5-valent and 7-valent cells) are observed on both types of curved structures. These defects, however, are absent in COS monolayers, and below we reanalyze the ascidian monolayers to detect the reason for this morphological difference. Namely, for the first time, we determine the values of $C'_{ij}$ and compare topological characteristics of the considered systems.

EMSG contained from 60 to 900 cells in the follicular system. The numbers of investigated samples for the studied species are: *Ciona intestinalis* — 21, *Molgula citrina* — 25, *Ascidiella aspersa* — 36, *Styela clava* — 14, *Styela plicata* — 8, *Ascidia mentula* — 11, *Molgula sp.* — 7. Topological analysis of typical samples of the weakly defective ascidian species *Styela clava* and more defective *Molgula sp.* is shown in panels (a-b) and (c-d), respectively. DOPTs for all investigated ascidian species are presented in panel (e).

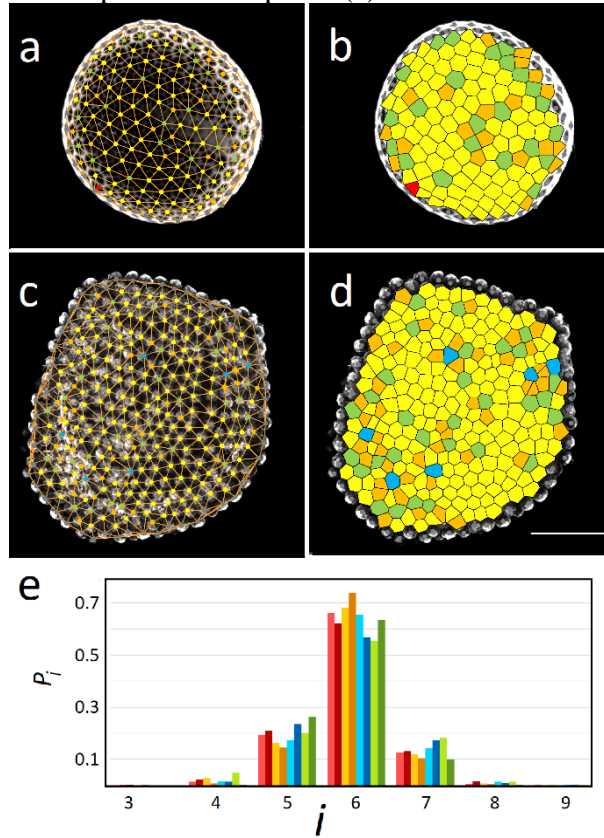

FIG. 3. Topological organization of epithelial follicular cells in ascidian species. (a,b) Typical image of *Styela clava* monolayer with superimposed Delaunay triangulation and Voronoi tessellation. (c,d) Typical sample of *Molgula sp.* monolayer and its similar graphical analysis. The color code is the same as in Fig. 2(a–d). Scale bar (common for a-d panels): 70 μm. (e) Distribution of cells according to the number of their sides for averaged data on ascidian monolayers of considered species. Red, dark red, orange, brown, blue, dark blue, green, dark green colors correspond to *Ascidiella aspersa, Molgula citrina, Styela clava, Ascidia mentula, Molgula sp., Phallusia mammillata, Styela plicata, Ciona intestinalis* monolayers, respectively.

Scars and pleats are observed in all types of ascidians, except for *Phallusia mammillata* and *Styela plicata* [11]. Thus, for scars and pleats to exist, the defectiveness of the order should be relatively low: $\tilde{Q} < 0.14$. In Fig. 3(b) (*Styela clava*) and (d) (*Molgula sp.*), the defects are highlighted as linear chains of orange and green polygons with 5 and 7 sides. Note that the last monolayer is transitional, and most of the topological defects are not linear but extended ones, although short pleats and scars are also observed. Thus, we conclude that in the spherical COS monolayers such linear topological defects are not possible due to the too large defectiveness: for the monolayers $\tilde{Q} \approx 0.27$.

The spherical geometry of the ascidian monolayers manifests itself in an essentially positive value of $\Delta$, see the next column of Table I. Pair correlations in the ascidian EMSG were calculated in the same way as in the COS monolayers. Below, let us make a few additional notes.

According to the calculated values of $<\tilde{Q}>$, the most ordered EMSG were observed in the species *Ciona intestinalis* and *Styela clava*. The *Ciona intestinalis* monolayers included relatively small numbers of cells with rather close sizes. About 48% of *Ciona intestinalis* one-side images had no extended topological defects (ETDs) at all. However, despite the highest ordering among ascidians, due to the too small size of the samples, we were able to determine $C'_{66}$ for monolayers of *Ciona intestinalis* only with low accuracy: the standard deviation is several times greater than the absolute value of $C'_{66}$. This occurs because the corresponding value $<N> = 94$ is small, and when processing images, it turns out that only a very small (15 on average) number of Voronoi cells have reliably determined boundaries. A bit more defective monolayers of the *Styela clava* species [see the example shown in Fig. 3(a,b)], despite the dominance of the hexagonal order, already have a certain number of scars and pleats. These monolayers have the relatively large correlation of $C'_{57}$ and the largest correlation $C'_{66}$ (see Table I).

TABLE I. Comparative analysis of different ascidian monolayers. The table lines are ordered according to the relative topological defectiveness $<\tilde{Q}>$ of monolayers (see the third column). This defectiveness is averaged over all the studied samples of one species; the cell number $<N>$ is averaged analogously. The value of $\Delta$ corresponds to the DOPTs shown in Fig. 3(e). The last six columns show averaged values of $C'_{ij}$ and their standard deviations for $i,j = 5,6,7$. The other correlations are not shown in the table because of the smallness of $P_4$ and $P_8$.

| Species | $<N>$ | $<\tilde{Q}>$ | $\Delta$ | $100 C'_{66}$ | $100 C'_{57}$ | $100 C'_{65}$ | $100 C'_{67}$ | $100 C'_{55}$ | $100 C'_{77}$ |
|---|---|---|---|---|---|---|---|---|---|
| *Ciona intestinalis* | 94 | 0.04 | 0.16 | -0.89±5.34 | 3.36±3.27 | 0.7±4.82 | 0.89±3.8 | -3.35±1.13 | -0.96±1.17 |
| *Styela clava* | 441 | 0.06 | 0.08 | 6.29±3.25 | 2.64±1.17 | -3.96±1.91 | -1.52±1.59 | -0.88±0.66 | -0.89±0.8 |
| *Ascidia mentula* | 615 | 0.084 | 0.04 | 3.52±1.66 | 3.29±0.88 | -2.73±1.39 | -2.53±1.43 | -0.9±0.41 | -0.46±0.45 |
| *Ascidiella aspersa* | 222.7 | 0.11 | 0.09 | 3.71±3.14 | 3.45±1.5 | -2.91±2.91 | -1.48±2.48 | -1.53±1.04 | -0.94±1.01 |
| *Molgula citrina* | 179.8 | 0.117 | 0.09 | 2.97±3.52 | 2.55±2.09 | -1.17±2.92 | -1.22±4.04 | 0.52±0.91 | 0.25±0.49 |
| *Molgula sp.* | 947 | 0.139 | 0.03 | 2.76±1.26 | 3.22±0.7 | -2.33±0.59 | -1.21±1.33 | 0.67±0.62 | 0.13±0.12 |
| *Phallusia mammillata* | 353 | 0.165 | 0.07 | 2.49±2.52 | 4.19±1.25 | -2.47±1.88 | -0.53±2.42 | 0.7±0.73 | 0±0.08 |
| *Styela plicata* | 404 | 0.18 | 0.08 | 4.82±2.37 | 3.06±0.88 | -3.18±1.96 | -0.25±1.36 | 0.58±0.43 | 0.12±0.2 |

On the surface of more defective monolayers ($\tilde{Q} \approx 0.14$) of the *Molgula sp.* samples, [see a typical example in Fig. 3(c,d)], extended defective areas are observed, where it is not possible to distinguish individual scars. Such mixing of the hexagonal order with scars leads to a decrease in the $C'_{66}$ value compared to the less defective ascidian species. The presence of extended topological defects and the observed increase in the length of scars and the number of pleats [see Fig. 3(c,d)] lead to an increase in the $C'_{57}$ value.

## V. MORPHOLOGY OF SPHERICAL MODEL PACKINGS

It is of interest to discuss (see the next section) to what extent our results obtained for the spherical epithelial monolayers can be reproduced within the framework of the simplest theoretical models. Let us first consider random spherical packings and packings obtained using the random sequential adsorption (RSA) method [27]. To obtain a random spherical packing, the points are randomly (with equal volume probability) successively ejected into a cube with side length of $d$. If the ejected point simultaneously locates into a sphere (with diameter $d$) inscribed in the cube, then its coordinates are projected onto the sphere surface along its radius. The process of ejection and selection of particles is repeated until the required number of points $N$ is reached. Then, the obtained points are used as nodes of the Voronoi tessellation. An example of the resulting packing (with $N = 300$) is shown in Fig. 4(a,e).

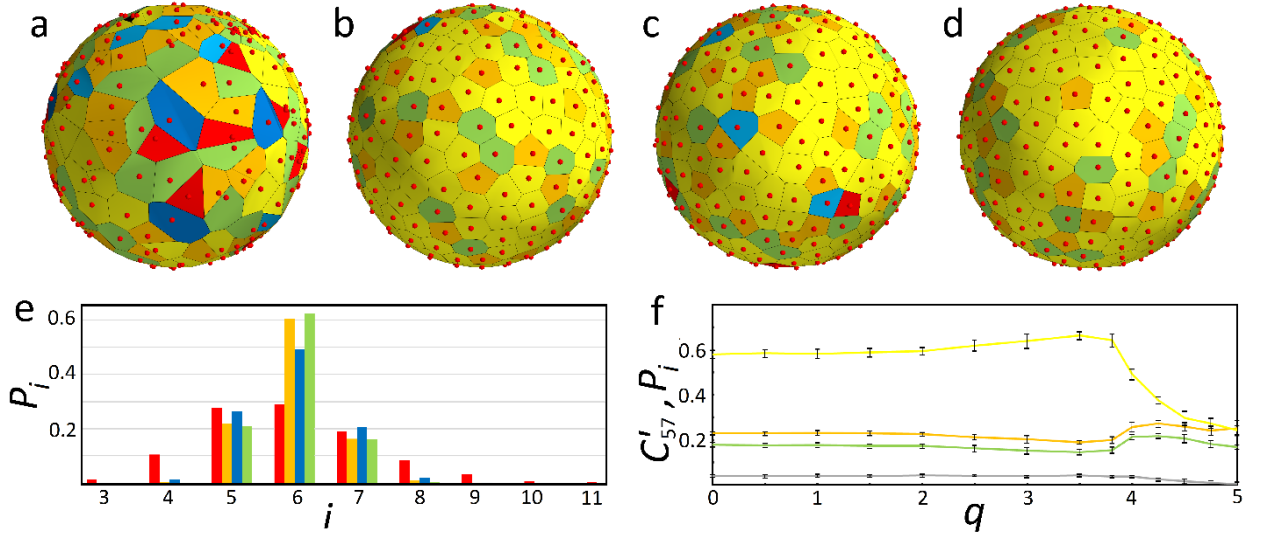

FIG. 4. Topological organization of model spherical structures containing $N = 300$ point particles. (a) Voronoi tessellation for a typical random spherical packing. (b) Structure resulting from (a) due to energy (7) minimization. (c) Voronoi tessellation for a typical spherical packing obtained with random sequential absorption method. (d) Structure resulting from (c) due to energy (7) minimization. The color code of the tessellations is the same as in Figs. 1 and 2. The cells with 3 and 9 neighbors (these cell valences are absent in Figs. 2 and 3) are shown with black and dark blue, respectively. (e) Distribution of cells according to the number of their sides for the averaged data on the spherical structures of four considered types. Red and dark blue colors correspond to the averaged data obtained for 10 random and 10 RSA spherical packings, while orange and green colors correspond to the averaged data obtained for the spherical structures that result from these random and RSA packings due to energy (7) minimization at the following parameters: $\beta S_{av}/\zeta = 1$ and $q = 3.81$. In terms of $P_6 \approx 0.49$, RSA packings are close to normal (invariant) proliferative epithelium, in which $P_6$ varies within 0.38-0.48 [2]. (f) Averaged dependencies of $P_5$, $P_6$, $P_7$ and $C'_{57}$ on $q$ for the energy-minimized structures (obtained from the RSA packings) are shown with orange, yellow, green and grey colors; for these plots $\beta S_{av}/\zeta = 1$, and vertical segments correspond to the values of standard deviations.

More ordered packings with a higher probability $P_6$ are the quasi-random ones obtained in the framework of the RSA method. Unlike random spherical packing, the points projected onto a sphere are considered to be the centers of balls with equal diameter $d$. If the ball corresponding to the last projected point do not overlap any previously adsorbed ball, the last point is accepted and remain fixed for the rest of the process. Otherwise, the last point and the corresponding ball are removed. The process stops when there are no more spaces left on the sphere surface to accommodate at least one more ball. Then the points (the centers of the balls) are considered as the centers of the cells of the Voronoi tessellation. An analysis of the topology of a typical sample of the RSA structure is shown in Fig. 4(c,e).

Obviously, there is an averaged maximum limit on the number of balls $N_{max}$ that can be placed on the sphere. This limit, also known as the jamming limit [37], has been extensively studied for various shapes and dimensions of adsorbed particles [38-40]. In the case of the random sequential adsorption of equivalent disks on the plane, the average ratio of the area occupied by them to the total surface area tends to $L \approx 0.547$ [37]. Obviously, in the spherical case this ratio tends to the same limit, if the ratio $D/d$, where $D$ is the sphere diameter, tends to infinity. Therefore, to generate spherical RSA packings with $\sim N$ particles, where $N$ is sufficiently large, the ratio $D/d$ should be approximately equal to $\sqrt{N/4L}$.

The obtained data for 10 random and 10 RSA spherical packings are shown in Table II. All studied model structures contained $N = 300$ Voronoi cells. Averaged DOPTs for these cases are shown in Fig. 4(e). The data presented in Table II show that random spherical packings are more disordered ones. They are characterized by a small fraction of cells with a hexagonal environment ($P_6 = 0.29$) and the absence of any pair correlations. According to their topological characteristics, these packings are far from the considered cell monolayers. More ordered RSA packings are characterized by the significant fraction of hexagonal cells ($P_6 = 0.49$), positive pair correlation $C'_{57}$ and negative correlations $C'_{55}$ and $C'_{77}$. The $C'_{66}$ correlation is close to zero. Thus, in terms of $P_6$ value, the RSA packings are somewhat more ordered than COS monolayers and less ordered than ascidian epithelium. In terms of pair correlations, these packings are close to COS monolayers.

TABLE II. Comparative analysis of different model structures. The table lines are ordered according to the relative topological defectiveness $<\tilde{Q}>$ (see the second column). The last six columns show averaged values of $C'_{ij}$ with their standard deviations for $i,j = 5,6,7$. The other correlations are not presented in the table because of the smallness of $P_4$ and $P_8$ probabilities.

| Model structure | $<\tilde{Q}>$ | $100 C'_{66}$ | $100 C'_{57}$ | $100 C'_{65}$ | $100 C'_{67}$ | $100 C'_{55}$ | $100 C'_{77}$ |
|---|---|---|---|---|---|---|---|
| Random | 0.51 | 0.04±0.65 | 0.6±0.57 | -0.1±1.05 | 0.21±0.71 | -0.57±0.39 | -0.26±0.57 |
| RSA | 0.26 | 0.02±0.85 | 4.00±0.85 | 0.44±1.08 | 0.01±0.70 | -2.57±0.42 | -1.86±0.47 |
| Minimized random | 0.19 | 0.47±1.22 | 3.49±0.61 | -0.32±0.6 | -0.13±1.09 | -1.98±0.27 | -1.71±0.49 |
| Minimized RSA | 0.17 | 0.19±1.03 | 3.71±0.52 | -0.02±1.02 | -0.31±1.14 | -1.93±0.43 | -1.68±0.39 |

Various approaches considering the elastic energy of cell monolayers can be found in the literature [13,27-29,41-43]. Here we use a well-known vertex model [27], in which cell centers and boundaries, like in our morphological analysis, are directly determined by the Voronoi tessellation. Within the framework of this model, it is possible to obtain more ordered spherical structures with DOPT similar to that observed in the non-proliferative spherical epithelium of ascidians.

When applying the model, we assume that the initial positions of the cell centers coincide with the centers of the Voronoi cells in the random spherical or RSA packings described above. Then, we minimize the elastic deformation energy $E$ with respect to the coordinates of cell centers using the coordinate descent method [44]. Following the works [28-29], the energy $E$ of a monolayer containing $N$ cells can be written as:

$$E = \sum_{i=1}^{N} \beta_i (A_i - A_i^0)^2 + \zeta_i (P_i - P_i^0)^2, \qquad (6)$$

where the subscript $i$ labels each cell; $A_i$ and $P_i$ are the cell area and perimeter, respectively. The origin of the elastic moduli $\beta_i$ is associated with a combination of the cell volume incompressibility and resistance to height differences between nearest cells [28-29]. The second term including the cell perimeters $P_i$ results from active contractility of the actomyosin subcellular cortex and effective cell membrane tension due to cell-cell adhesion and cortical tension, which are linear in perimeter. Usually, different cells of the same type are assumed to have the same elastic moduli $\beta_i$ and $\zeta_i$ and, thus, index '$i$' is omitted. The same goes for the parameters $P_i^0$ and $A_i^0$, which are usually substituted with $P_{eff}$ and $S_{av}$, respectively. The resulting energy reads:

$$E = \sum_{i=1}^{N} \beta (A_i - S_{av})^2 + \zeta (P_i - P_{eff})^2. \qquad (7)$$

The value of $S_{av}$ is equal to the ratio between the area of the sphere and the number of cells. As is known, the mechanical properties of the monolayer described by energy (7) are determined by the target shape index $q = P_{eff}/\sqrt{S_{av}}$ [28,29]. In the model planar monolayer, the transition between glass-like state ($q < q_0$) and fluid-like one ($q > q_0$) occurs at the point $q_0 \approx 3.81$ of the so-called jamming transition [28,29]. In the latter state the monolayer is less ordered since its cells take more elongated shape, and the fraction of 6-valent cells decreases [28,29]. Below, following [28,29] we consider morphological properties of spherical model structures in dependence on $q$ (instead of $P_{eff}$). The 10 random and 10 RSA spherical packings (already mentioned above) were used as the initial ones for the subsequent minimization of energy (7).

Let us begin from the case when two terms in the energy (7) are comparable and, consequently, $\beta S_{av}/\zeta = 1$. Then our analysis shows that the most ordered structures (with the highest values of $P_6 \gtrsim 0.6$) appear if $q = 3.81$ or a bit smaller. Figure 4 and the last two rows of Table 2 demonstrate averaged data for spherical model structures with $q = 3.81$. According to our computations, when $q$ increases, the value of $P_6$ rapidly decreases: at $q = 4.5$ this probability achieves ~0.3, which is typical for random packing. In contrast, when q decreases even up to zero, only a slight decrease (5-10%) in the value of $P_6$ occurs. Also note that in all the region $q < 3.5$ the value $C'_{57} \approx 0.04$ (see Fig. 4f).

As we have also revealed, a change in the value of $\beta S_{av}/\zeta$ in the range from 0.1 to 10 has practically no effect on the topological characteristics of the model epithelium. The dependence of these characteristics on the initial coordinates of the nodes is also rather weak. In the limiting cases, when $\beta = 0$ or $\zeta = 0$, the probability $P_6$ decreases by about 10% at the point $q = 3.81$, which is apparently because the most ordered packing is that in which both the areas and perimeters of cells simultaneously tend to their average values. In the case of $\beta = 0$ one can consider the $q$ value variation. If $q$ decreases and is smaller than 3.81, an additional slight decrease (up to 5-10%) in the value of $P_6$ occurs. When $q$ increases and is greater than 3.81, the value of $P_6$ rapidly decreases in the same way as in the case of $\beta S_{av}/\zeta = 1$.

Summing up, we conclude that the DOPTs of the model spherical packings considered in the region 3.5<$q \leq 3.81$ and $0.1 < \beta S_{av}/\zeta < 10$ are close to those observed in the non-proliferative epithelia of ascidians. In terms of pair correlations, the model packings are different from the epithelium of ascidians only by a much smaller value of $C'_{66}$. However, this fact is probably explained by the presence of gap junctions [45] in the ascidian epithelial follicular system (see more details in the next section).

For the first glance, in the region $q > 3.81$ the vertex model could explain the topology of proliferative COS monolayers, which are more disordered than the epithelium of ascidians (compare DOPTs in Fig. 3e and Fig. 2e). However, such application of the model is not physical, since, when $q$ increases, the monolayer transforms into a liquid-like state. To make the glass-like state monolayer (at $q < 3.81$) more disordered, one could try to consider a disordering induced by cell division and dynamics (which are substantially important for MDCK monolayers [46,47]), but this study is beyond the scope of our work.

Using periodic boundary conditions, we analogously considered planar model monolayers also containing $N = 300$ cells in the fundamental region. These boundary conditions lead to the exact fulfillment of the equality $\Delta = 0$ and DOPT in the planar monolayers becomes balanced: in the planar monolayers, compared with the spherical ones, the probability of $P_5$ decreases by about 0.04, and $P_6$ increases by the same amount. However, we have not found any other significant peculiarities, which are not explained by random spread in the model structures.

Concluding the consideration of structures obtained within the vertex model we note they are characterized by large values of $P_6$ (close to ~0.6) and values of pair correlations similar to those obtained for RSA packings. The topological characteristics of the energy-minimized structures are practically independent on the initial order of the packings (compare the last two rows of Table II).

The number of random and RSA structures considered in this work is relatively small, but quite acceptable for calculating DOPTs [see Fig. 4(e)]. One can justify this point by referring to the fact that the DOPT obtained by us for the random spherical packing is very close to the data [44] obtained for the planar case. The values of pair correlations obtained are obviously very approximate, but they indicate the sign or closeness to zero of the considered correlations. This is sufficient for the qualitative comparative analysis of the morphology of the structures considered in the article, which is carried out in the next section.

## VI. DISCUSSION AND CONCLUSION

In this study that continues series of our papers [10,11,13], we analyzed and compared the topological characteristics of the cell order in several monolayer systems. Considering spherical and planar COS cell monolayers, we have shown that the spherical shape of the epithelium results in an imbalance of DOPT histograms, numerically described by the quantity $\Delta \neq 0$ [see Eq. (1)]. Namely, in contrast to the case of planar epithelium, the right-side sum calculated for the histogram $\sum_{i \geq 7} P_i \cdot (i - 6)$ turns out to be slightly larger than the left-side one $\sum_{i \leq 5} P_i \cdot (6 - i)$. Since the main contribution to these two expressions is made by the terms with $P_5$ and $P_7$, the fraction of 5-valent cells $P_5$ in COS monolayers is greater in the spherical case, than in plane one, and the fraction of 7-valent cells $P_7$ is smaller in the plane epithelium. As we have shown, the imbalance is especially noticeable for a spherical epithelium containing several hundred of cells. A decrease in the value of $\Delta$ for ascidian monolayers of *Molgula sp. and Ascidia mentula* (see Table I) takes place because the value is inversely proportional to the number of cells in a spherical structure and the *Molgula sp. and Ascidia mentula* samples are characterized by the largest cell numbers $< N >$.

The fact that the linear topological defects typical of colloidal crystals appear in the spherical epithelium of ascidians was revealed in our previous work [11]. In addition, it has been shown that due to the greater spread of cell sizes, scars and pleats in spherical epithelium usually appear in the structure with $N = 200 - 300$, while for formation of scars and pleats in colloidal crystals $N$ should exceed ~400 [34]. However, in spherical COS cell monolayers, which have approximately the same number of cells (several hundreds) as monolayers of some species of ascidians, linear topological defects are practically not observed. As we have found out in this work, this is due to a high degree of disorder in the proliferative COS monolayers. Note that the quantity $\sum_{i \geq 7} P_i \cdot (i - 6)$, introduced above during the discussion of the properties of DOPT, is also the measure $\tilde{Q}$ of the topological defectiveness of the spherical monolayers. Based on the analysis of the data on ascidian monolayers, we have found the critical value $\tilde{Q} \sim 0.14$, above which the linear topological defects are almost not observed due to a high degree of topological disorder. This is also true for COS cell monolayers, where $< \tilde{Q} > \sim 0.27$.

We also studied the regularities in the locations of nearest cells depending on their valency and the monolayer geometry. Considering a virtual random polygonal packing with the same DOPT as in the real structure, we determine the nonrandom contribution $C'_{ij}$ to the pair correlation of neighboring $i$- and $j$-valent cells. The positive/negative value of $C'_{ij}$ reveals the tendency to nonrandom agglomeration/repelling of nearest cells with the corresponding valences. These correlations can be significant provided that the values $P_i$ are large enough; this is true for $i,j$=5,6,7. Also, we note that for the considered planar and spherical structures, the correlation sign is perfectly explained within the concept of topological charge, which equals to $6 - i$ for a single $i$-valent cell. For all examples of the packings considered in this paper, nonrandom correlations can be essentially positive if and only if the corresponding pair of nearest cells has zero total topological charge. This fact recalls the results of the famous Berezinskii–Kosterlitz–Thouless theory [48,49], in which vortices and antivortices with opposite topological charges are attracted to each other.

The main common morphological feature of both considered biosystems, RSA and energy minimized spherical structures is the presence of an essential positive correlation $C'_{57}$, the value of which exceeds the standard deviation of the corresponding averagings [see Tables I-II and Fig. 2(g)]. In sufficiently ordered spherical monolayers of ascidians (corresponding to rows 2–5 of Table I), one can trace the relationship between the $C'_{57}$ value and the frequency of occurrence and the length of linear topological defects. It is quite natural since in these defects 5- and 7-valent cells alternate. However, these same correlations are also observed in various planar and spherical structures (even without scars and pleats), and we could not detect any specific relationship between the shape of the monolayer and the magnitude of this (or any other) pair correlation.

The main difference between the two considered biosystems is that all ascidian species are characterized by a higher value of $C'_{66}$ correlation and, accordingly, in the non-proliferative epithelium of ascidians, 6-valent cells tend to agglomerate. In our opinion, this fact can be explained by the presence of gap junctions [45] in the ascidian epithelial follicular system. These junctional complexes, absent in COS monolayers, are intercellular channels connecting adjacent cells electrically and metabolically providing a better coordination of cellular functions in all metazoans. In this physiological context, it may be beneficial for cells to have approximately the same cell side lengths, which is best realized in agglomerations of 6-valent cells.

Finally, we hypothesize that in normal biological tissues, specific structural peculiarities including the invariant distribution of cell valences, specific topological defects and certain pair correlations generate a subtle relation between order and disorder determining the epithelial tissue elasticity and authorizing the stable and reproducible deformations of mono-stratified epithelia throughout the morphogenetic processes. In the near future, we hope to further justify this assumption by analyzing the relations between order, disorder and epithelial deformations during zebrafish embryonic development in vivo.

## APPENDIX A: CELL CULTURE

COS cells (COS-1 (ATCC CRL-1650), gift from the "Institut de Biothérapie", Montpellier, France) were seeded either on standard Petri dishes (flat monolayers), or, to induce spherical monolayers, on hydrophobic Petri dishes in MEM alpha medium supplemented with 10% fetal bovine serum, 20 ng/ml HGF, 10 ng/ml EGF, 25 mM glucose, 1 μMthyrotropin-releasing hormone, 1 μMhydrocortisone, 10 μg/ml insulin, 50 μg/ml albumin-linoleic acid, 0.1 μMselenium acetate, 0.5 μg/ml ferrous sulfate, 0.75 μg/ml zinc sulfate, 10 mM nicotinamide, 100 μg/ml streptomycin and 100 U/ml penicillin.

## APPENDIX B: EGG COLLECTION

Ascidians were collected in Roscoff (Bretagne Nord, France, latitude: 48.726199, longitude: -3.985324999999989) and their oocytes maintained at 18 °C in 0.2 μm-filtered seawater containing 100 U/ml penicillin, and 100 μg/ml streptomycin. Oogenesis occurs continuously throughout adulthood [50, 51] and, oocytes, at different stages of folliculogenesis [18, 52], were obtained through gonad dislocation. Experiments conducted on ascidians (marine invertebrates) did not require specific permissions and were collected outside of private or protected area. Our field studies did not involve endangered or protected species.

## APPENDIX C: LABELING AND FLUORESCENCE MICROSCOPY

COS cells and ascidian oocytes were fixed and processed for cell contour (TRITC phalloïdin labeling) and nuclei (DAPI labelling) as previously described [53, 54]. Specimens were analyzed with a Leica SPE laser confocal microscope (Montpellier RIO Imaging platform, France).

## APPENDIX D: IMAGE ANALYSIS

After determining the geometric centers of the cell nuclei, triangulation was performed using the Delaunay method [32]. Next, the Voronoi tiling was constructed, and the areas of the epithelial cells were calculated as the areas of Voronoi cells. Obviously, for a correct statistical analysis, it is necessary to discard the cells (located too close to the image border), the number of neighbors for which cannot be determined. Note that even if it is possible to construct a closed Voronoi cell, then it is also necessary to check whether the cell polygon boundary can be changed by additional hypothetical nuclei lying directly outside the image border. Therefore, the center of a reliably constructed Voronoi cell should be located at least twice as far from the image border as any of the vertices of this cell.

# APPENDIX E: CALCULATING PAIR CORRELATIONS FOR A SMALL FRAGMENT OF DELAUNAY TRIANGULATION

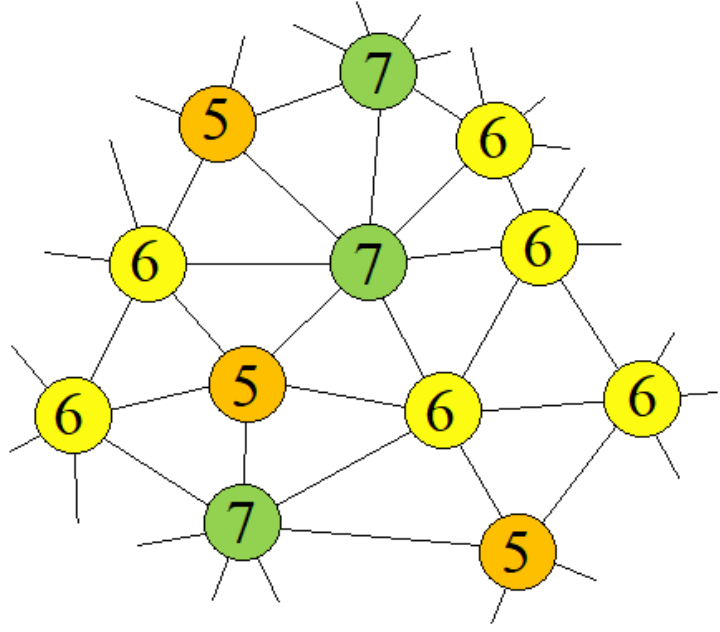

FIG. 5. Toy fragment of Delaunay triangulation. The color of a circle located at a vertex of the triangulation corresponds to the color code of Figs. 2-4 and is determined by the number of edges outgoing from the vertex.

Six of the 12 vertices present in the fragment (see Fig. 5) have 6 neighbors, so $P_6 = \frac{6}{12} = 0.5$. Similarly, $P_5 = P_7 = \frac{1}{4} = 0.25$. Now let's calculate the pair correlations $C_{ij}$ for the considered fragment. To do this, we consider only those triangulation edges that fall entirely into the fragment under consideration. There are 22 such edges, therefore $L_{tot} = 22$. Six edges connect vertices that have 6 neighbors, so $L_{66} = 5$ and $C_{66} = \frac{L_{66}}{L_{tot}} = \frac{5}{22} \approx 0.23$. Similarly, $C_{57} = \frac{L_{57}}{L_{tot}} = \frac{5}{22} \approx 0.23$.

Using the formula $\sum_k (NP_k k) = 12 \cdot (0.25 \cdot 5 + 0.5 \cdot 6 + 0.25 \cdot 7) = 72$ (see Eq. (2)) one can find exactly the number of ends of the edges shown in the figure. Indeed, 24 fully shown edges have 48 ends, and 24 edges go beyond the borders of the picture.

Let's calculate the fractions of edges in hypothetical (random) triangulated structure that is characterized by the same DOPT. Using Eq. (3) the fraction $P_{5-7}^{rnd}$ of triangulation edges connecting 5-valent and 7-valent cells can be calculated: $P_{57}^{rnd} = \frac{2 \cdot 5 \cdot 7 \cdot P_5 \cdot P_7}{[5P_5 + 6P_6 + 7P_7]^2} = 0.12$. Similarly, using formula (4), one can find that $P_{66}^{rnd} = \frac{36P_6^2}{[5P_5 + 6P_6 + 7P_7]^2} = 0.25$. Substituting these values into formula (5), one can obtain that the nonrandom contributions to the pair correlation function are equal to $C'_{57} = C_{57} - P_{57}^{rnd} = 0.11$; $C'_{66} = C_{66} - P_{66}^{rnd} = -0.02$. Thus, in this fragment there is a negative correlation $C'_{66}$ and positive one $C'_{57}$. In fact, the positive correlation $C'_{57}$ occurs since in this structure all 5- and 7-valent cells are not distributed uniformly over the triangulation fragment under consideration, but are combined into one scar, which is a topological defect, where 5- and 7-valent particles alternate.

## AUTHOR CONTRIBUTIONS

## ACKNOWLEDGEMENT

D.R. and K.F. acknowledge financial support from the Russian Science Foundation, grant No. 22-72-00128. The authors thank A.E. Myasnikova for fruitful discussions.